\documentstyle[prd,aps,epsfig]{revtex}
\begin{document}
\title
{Exclusive semileptonic decays of heavy mesons in quark model}
\author{Dmitri Melikhov}
\address{Nuclear Physics Institute, Moscow State University, Moscow, 119899, Russia
\thanks{Electronic address: melikhov@monet.npi.msu.su}}
\maketitle
\begin{abstract} 
Semileptonic decays $D\to K,K^*,\pi,\rho$, $D_s\to \eta,\eta',\phi$, 
and $B\to D,D^*,\pi,\rho$ are analyzed within the dispersion formulation of the quark 
model. The form factors at timelike $q$ are derived by the analytical 
continuation from spacelike $q$ in the form factors of the relativistic light--cone 
quark model. The resulting double spectral representations allow a direct calculation of the 
form factors in the timelike region. The results of the model are shown to be in good 
agreement with all available experimental data. From the analysis of the $B\to D,D^*$ decays we find 
$|V_{cb}|=0.037\pm 0.004$, and the $B\to\pi,\rho$ decays give $|V_{ub}|=0.004\pm 0.001$. 
\end{abstract}
\vspace{0.5cm}
%
The interest in exclusive semileptonic decays of heavy mesons lies in a possibility 
to obtain the most accurate values of the quark mixing angles and test various
approaches to the description of the internal hadron structure. The decay rates are 
expressed through the Cabbibo--Kobayashi--Maskawa (CKM) matrix elements 
and hadronic form factors of the weak currents which contain the 
information on hadron structure. 
New accurate data on $B\to D,D^*$ and first 
measurements of the $B\to\pi,\rho$ decays open a possibility 
to determine $V_{cb}$ and $V_{ub}$ with high accuracy and require reliable theoretical 
predictions on the form factors and decay rates. 
A nonperturbative theoretical study 
should give these form factors in the whole kinematical region of momentum transfers 
$0\le q^2 \le (M_1-M_2)^2$, $M_1$ and $M_2$ the initial and final meson masses, respectively.

The amplitudes of meson decays induced by the quark transition $q_i\to q_f$ through 
the vector $V_{\mu}\;=\;{\bar q_f}\gamma_{\mu}q_i$ and 
axial--vector $A_{\mu}\;=\;{\bar q_f}\gamma_{\mu}\gamma_{5}q_i$ currents 
have the following structure
\cite{iw} 
\begin{eqnarray}
\label{amplitudes}
<P(M_2,p_2)|V_\mu(0)|P(M_1,p_1)>&=&f_+(q^2)P_{\mu}+f_-(q^2)q_{\mu},  \nonumber \\
<V(M_2,p_2,\epsilon)|V_\mu(0)|P(M_1,p_1)>&=&2g(q^2)\epsilon_{\mu\nu\alpha\beta}
\epsilon^{*\nu}\,p_1^{\alpha}\,p_2^{\beta}, \nonumber \\
<V(M_2,p_2,\epsilon)|A_\mu(0)|P(M_1,p_1)>&=&
i\epsilon^{*\alpha}\,[\,f(q^2)g_{\mu\alpha}+a_+(q^2)p_{1\alpha}P_{\mu}+
               a_-(q^2)p_{1\alpha}q_{\mu}\,],  
\end{eqnarray}
with $q=p_{1}-p_{2}$, $P=p_{1}+p_{2}$.
We use the notations:  
$\gamma^{5}=i\gamma^{0}\gamma^{1}\gamma^{2}\gamma^{3}$, 
$\epsilon^{0123}=-1$,  
$Sp(\gamma^{5}\gamma^{\mu}\gamma^{\nu}\gamma^{\alpha}\gamma^{\beta})=
4i\epsilon^{\mu\nu\alpha\beta}$.
In general all the form factors are independent functions to be calculated within a 
nonperturbative approach. 
Considerable simplifications occur when both the parent and the daughter quarks active 
in the weak transition are heavy. 
Due to the heavy quark symmetry (HQS) \cite{iw} all the form factors which are in general 
functions of $q^2$ and the masses,  
depend on the one dimensionless variable $\omega=p_1p_2/M_1M_2$. 
It is convenient to introduce dimensionless 'heavy quark' form factors 
\begin{eqnarray}
\label{hh}
&f_\pm(q^2)=\frac{M_2\pm M_1}{2\sqrt{M_1M_2}}h_+(\omega)
+\frac{M_2\mp M_1}{2\sqrt{M_1M_2}}h_-(\omega),
\nonumber \\
&g(q^2)=\frac{1}{2\sqrt{M_1M_2}}h_g(\omega),\quad
a_+(q^2)=-\frac{1}{2\sqrt{M_1M_2}}h_a(\omega),\quad
f(q^2)=\sqrt{M_1M_2}(1+\omega)h_f(\omega). 
\end{eqnarray}
In the leading $1/m_Q$ order all the form factors $h$ 
can be expressed through the single universal form factor, the Isgur--Wise function $\xi$ as 
follows 
\begin{eqnarray}
\label{hhiw}
h_+(\omega)=h_g(\omega)=h_a(\omega)=h_f(\omega)=\xi(w), \qquad h_-(\omega)=0.
\end{eqnarray}
The normalization of the Isgur--Wise function at zero recoil is known, $\xi(1)=1$. 
In contrast to meson decays induced by the heavy--to--heavy quark transitions, 
the general case of the transitions 
between hadrons with arbitrary masses, and in particular meson decays induced by a 
heavy--to--light quark transitions are not well--understood.
To date, theoretical predictions on semileptonic decays 
induced by heavy--to--light quark transitions coming from 
the quark model 
\cite{wsb,isgw,gs,jaus,faustov,simula,cheng}, 
QCD sum rules \cite{bbd,ball1,narison,kr,ball2}, and lattice calculations
\cite{lat1,lat2,ape,lat4} 
differ significantly. 
Recently, B.~Stech noticed \cite{stech} that new relations between the form 
factors of meson transition can be derived if use is made of the constituent quark
picture. These relations are based on the observation that if the meson wave function in terms of its 
quark constituents is strongly peaked in the momentum space with the width of order 
$\Lambda\simeq 0.5\; GeV$ then for a heavy parent quark  
a small parameter $\Lambda/m_Q$ appears in the picture and one can derive the
leading--order expressions for the form factors of interest which turn out to be
independent of subtle details of the meson structure. 
Although these relations give a guideline for the analysis of the decay processes, 
they cannot substitute calculations of the form factors in a more detailed dynamical model. 
The kinematically accessible $q^2$--interval in the meson decay induced by the 
heavy--to--light transition is $O(m_Q^2)$ so a relativistic treatment is necessary. 

A relativistic light--cone quark model (LCQM) \cite{jaus} is an adequate framework for 
considering decay processes. The model is formulated at spacelike momentum transfers and 
the direct application of the model at timelike 
momentum transfer is hampered by pair--creation subprocesses which at $q^2>0$ cannot be killed 
by an appropriate choice of the reference frame.
In \cite{jaus} the form factors at $q^2>0$ were obtained by numerical extrapolation from the 
region $q^2<0$. As the relevant $q^2$--interval is large, the accuracy of such a procedure is not
high. In \cite{simula,cheng} the nonpartonic contribution of pair--creation subprocesses is neglected and 
only the partonic part of the form factor is calculated in the whole kinematical interval of $q^2$.
Unfortunately, the nonpartonic contribution is under control only at $q^2=0$ where it vanishes. 

The dispersion formulation of the LCQM proposed in \cite{m1} overcomes these difficulties as it 
allows the analytical continuation to the timelike $q$. We apply this approach
to calculating the transition form factors of the semileptonic decays. 
 
The transition of the initial meson  $Q(m_2)\bar Q(m_3)$ with the mass $M_1$ 
to the final meson $Q(m_2)\bar Q(m_3)$ with the mass $M_2$ 
induced by the quark transition $m_2\to m_1$ is given by the diagram of Fig.1. 
The constituent quark structure of the initial and final mesons are described by the vertices $\Gamma_1$
and $\Gamma_2$, respectively. The initial pseudoscalar meson the vertex has the spinorial structure
$\Gamma_1=i\gamma_5G_1/\sqrt{N_c}$; the final meson vertex has the structure 
$\Gamma_2=i\gamma_5G_2/\sqrt{N_c}$ for a pseudoscalar state and the structure 
$\Gamma_{2\mu}=[A\gamma_\mu+B(k_1-k_3)_\mu]\,G_2/\sqrt{N_c}$ for the vector state. The values 
$A=-1$, $B=1/(\sqrt{s_2}+m_1+m_3)$ correspond to an $S$--wave vector meson, and 
$A=1/\sqrt{2}$, $B=[2\sqrt{s_2}+m_1+m_3]/[\sqrt{2}(s_2+(m_1+m_3)^2)]$ 
correspond to a $D$--wave vector particle. 

At $q^2<0$ the form factors of the LCQM \cite{jaus} can be represented 
as double spectral representations over the invariant masses of the initial and final
$q\bar q$ pairs as follows \cite{m1}
\begin{equation}
\label{ff}
f_i(q^2)=f_{21}(q^2)
\int\limits^\infty_{(m_1+m_3)^2}\frac{ds_2G_{2}(s_2)}{\pi(s_2-M_2^2)}
\int\limits^{s_1^+(s_2,q^2)}_{s_1^-(s_2,q^2)}\frac{ds_1G_{1}(s_1)}{\pi(s_1-M_1^2)}
\frac{\tilde f_i(s_1,s_2,q^2)}{16\lambda^{1/2}(s_1,s_2,q^2)},
\end{equation}
where 
$$
s_1^\pm(s_2,q^2)=
\frac{s_2(m_1^2+m_2^2-q^2)+q^2(m_1^2+m_3^2)-(m_1^2-m_2^2)(m_1^2-m_3^2)}{2m_1^2}
\pm\frac{\lambda^{1/2}(s_2,m_3^2,m_1^2)\lambda^{1/2}(q^2,m_1^2,m_2^2)}{2m_1^2}
$$
and 
$
\lambda(s_1,s_2,s_3)=(s_1+s_2-s_3)^2-4s_1s_2
$
is the triangle function. Here
$f_{21}(q^2)$ is the form factor of the constituent quark transition $m_2\to m_1$. 
In what follows we set $f_{21}(q^2)=1$. 

The double spectral densities $\tilde f_i(s_1,s_2,q^2)$ of the form factors have the 
following form: 
\begin{eqnarray}
\label{fplus}
&&\tilde f_++\tilde f_-=4\;[m_1m_2\alpha_1-m_2m_3\alpha_1+m_1m_3(1-\alpha_1)-m_3^2(1-\alpha_1)
+\alpha_2s_2],
\\
\label{fminus}
&&\tilde f_+-\tilde f_-=4\;[m_1m_2\alpha_2-m_1m_3\alpha_2+m_2m_3(1-\alpha_2)-m_3^2(1-\alpha_2)
+\alpha_1s_1],
\\
\label{g}
&&\tilde g=-2A\,[m_1\alpha_{2}+m_2\alpha_{1}+m_3(1-\alpha_{1}-\alpha_{2})]-4B\beta,
\\
\label{a1}
&&\tilde a_++\tilde a_-=-4A\,[2m_2\alpha_{11}+2m_3(\alpha_1-\alpha_{11})]+4B[C_1\alpha_1+C_3\alpha_{11}],
\\
\label{a2}
&&\tilde a_+-\tilde a_-=
-4A\,[-m_1\alpha_{2}-m_2(\alpha_{1}-2\alpha_{12})-m_3(1-\alpha_1-\alpha_2+2\alpha_{12})]
+4B\,[C_2\alpha_1+C_3\alpha_{12}],
\\
\label{flc}
&&\tilde f^{LC}=\frac{M_2}{\sqrt{s_2}}\tilde f_{D}+
\left({\frac{s_1-s_2-s_3}{2\sqrt{s_2}}-\frac{M_1^2-M_2^2-s_3}{2M_2}}\right)M_2\tilde a_+,
\end{eqnarray}
where 
\begin{eqnarray}
\label{fdisp}
\tilde f_D&=&-4A[m_1m_2m_3+\frac{m_2}2(s_2-m_1^2-m_3^2)
+\frac{m_1}2(s_1-m_2^2-m_3^2)-\frac{m_3}2(s_3-m_1^2-m_2^2)\nonumber\\
&&+2\beta(m_2-m_3)]+4B\,C_3\beta,
\end{eqnarray}

\begin{eqnarray}
\label{alpha1}
&\alpha_1=\left[(s_1+s_2-s_3)(s_2-m_1^2+m_3^2)-2s_2(s_1-m_2^2+m_3^2)\right]/{\lambda(s_1,s_2,s_3)},
\\
\label{alpha2}
&\alpha_2=
\left[(s_1+s_2-s_3)(s_1-m_2^2+m_3^2)-2s_1(s_2-m_1^2+m_3^2)\right]/{\lambda(s_1,s_2,s_3)},
\\
\label{beta}
&\beta=\frac14\left[2m_3^2-\alpha_1(s_1-m_2^2+m_3^2)-\alpha_2(s_2-m_1^2+m_3^2)\right],
\\
\label{alpha11}
&\alpha_{11}=\alpha_1^2+4\beta {s_2}/{\lambda(s_1,s_2,s_3)}, \quad
\alpha_{12}=\alpha_1\alpha_2-2\beta(s_1+s_2-s_3)/\lambda(s_1,s_2,s_3),
\\
\label{cc}
&C_1=s_2-(m_1+m_3)^2, \quad C_2=s_1-(m_2-m_3)^2, \quad C_3=s_3-(m_1+m_2)^2-C_1-C_2.
\end{eqnarray}

Let us underline that the representation (\ref{ff}) with the spectral densities
(\ref{fplus}--\ref{flc}) are just 
the dispersion form of the corresponding light--cone expressions from \cite{jaus}. 
It is important that double spectral representations without subtractions are 
valid for all the form factors except $f$ which requires subtractions. 
In the LCQM the particular form of such a representation for the form factor $f$
depends on the choice of the current component used for its determination and cannot be fixed
uniquely. In \cite{mn} the behaviour of the form factors of the vector, axial--vector and tensor
current has been studied in the limits of heavy--to--heavy and heavy--to--light quark transitions. 
The analysis of the behavior of the form factor $f$ in the case of a heavy--to--light quark
transition suggests another expression \cite{mn}: 
\begin{equation}
\label{fhq}
\tilde f^{HQ}(s_1,s_2,q^2)=\tilde f_D(s_1,s_2,q^2)+(M_1^2-s_1+M_2^2-s_2)\tilde g(s_1,s_2,q^2).
\end{equation}
We shall use both of these prescriptions in the numerical analysis of semileptonic decays. 
 
For a pseudoscalar or vector meson with the mass $M$ built up of the 
constituent quarks $m_q$ and $m_{\bar q}$, the function $G$ is normalized as 
follows \cite{m1}
\begin{equation}
\label{norma}
\int\frac{G^2(s)ds}{\pi(s-M^2)^2}\frac{\lambda^{1/2}(s,m_q^2,m_{\bar q}^2)}{8\pi s}
(s-(m_q-m_{\bar q})^2)=1. 
\end{equation}
The quark--meson vertex $G$ can be written as \cite{jaus}
\begin{equation}
\label{5vertex}
G(s)=\frac{\pi}{\sqrt{2}}\frac{\sqrt{s^2-(m_1^2-m_2^2)^2}}
{\sqrt{s-(m_1-m_2)^2}}\frac{s-M^2}{s^{3/4}}w(k),\qquad
k=\frac{\lambda^{1/2}(s,m_1^2,m_2^2)}{2\sqrt{s}}
\end{equation}
where $w(k)$ is the ground--state $S$--wave radial wave function. 

As the analytical continuation of the form factors (\ref{ff}) to the timelike region is performed, 
in addition to the normal contribution which is just the expression (\ref{ff}) taken
at $q^2>0$ the anomalous contribution emerges. The corresponding expression is given in
\cite{m1}. The normal contribution dominates the form factor at small timelike $q$ 
and vanishes as $q^2=(m_2-m_1)^2$ while the anomalous contribution is negligible at small $q^2$ and steeply rises
as $q^2\to(m_2-m_1)^2$. It should be emphasized that we derive the analytical continuation in the region
$q^2\le(m_2-m_1)^2$. For the constituent quark masses used in the quark models this allows a direct
calculation of the form factors of the $P\to V$ transitions in the whole kinematical decay region
$0\le q^2\le (M_P-M_V)^2$, as $M_P-M_V<m_2-m_1$. For the $P\to P'$ transition this is not the case: 
normally, $M_P-M_{P'}>m_2-m_1$. For the $P\to P'$ decays we directly calculate the form factors in the
region $0\le q^2\le (m_2-m_1)^2$ and perform numerical extrapolation in $(m_2-m_1)^2\le q^2\le
(M_P-M_{P'})^2$\footnote{The analytical continuation to $q^2>(m_2-m_1)^2$ is also possible. However one
should be careful when applying the constituent quark model for such $q^2$: we approach the unphysical 
$q\bar q$ threshold $q^2=(m_1+m_2)^2$ which is obviously absent in the amplitudes of hadronic processes.
This is a sign that we are coming to the region where the constituent quark picture is not adequate.}.
Numerical analysis shows the accuracy of this extrapolation procedure to be very high. 
We would like to notice that the direct calculation shows the derivative of the form factor $f_+$ to be
positive at the point $q^2=(m_2-m_1)^2$. This suggests that the maximum of the form factor 
$f_+$ at $q^2=(m_2-m_1)^2$ observed in \cite{simula,cheng} is just an artifact of neglecting the nonpartonic contribution to
the form factor. 

For calculating the form factors of semileptonic decays we assume that the wave function $w$ can be
approximated by a simple exponential function  
$w(k)=\exp(-k^2/2\beta^2)$
and adopt the numerical parameters of the 
ISGW2 model \cite{isgw2} shown in Table \ref{table:parameters}. 

The results of calculating the form factors are fitted by the functions 
$$
f(q^2)=f(0)/[1-\alpha_1q^2+\alpha_2q^4]
$$
with better than $0.5\%$ accuracy, and for the form factor $f_+$ this formula is used for numerical
extrapolation to the region $(m_1-m_2)^2\le q^2\le(M_1-M_2)^2$. 
The decay rates are calculated from the form factors via the formulas from \cite{gs}.  
Decay rate calculations are performed using the two prescriptions for the form factor $f$ given by the 
relations (\ref{flc}) and (\ref{fhq}); the corresponding results are 
labelled as LC and HQ, respectively. \\
{\bf The decay \boldmath$D\to K,K^*$\unboldmath}. \\
These CKM--favoured decays extend the widest possibility for detailed verification of the model. 
The parameters of the fits to the form factors are given in Table \ref{table:ffsd2k}. 
Using the value $V_{cs}=0.975$ \cite{pdg} the decay rates are found to be 
\begin{eqnarray}
\label{d2k}
\Gamma(D\to K)&=&8.7 \times 10^{10}\,s^{-1} \nonumber\\
\Gamma(D\to K^*)&=&
\left\{
\begin{array}{lll} 
5.58 \times 10^{10}\,s^{-1},\quad& \Gamma_L/\Gamma_T=1.34   \quad & (LC)\\
5.38 \times 10^{10}\,s^{-1},     & \Gamma_L/\Gamma_T=1.34         & (HQ)
\end{array}
\right.
\end{eqnarray}
Table \ref{table:ratesd2k} compares the results of the model with the $HQ$ prescription for the form
factor $f$ with the experimental data. One can observe perfect agreement with the data.
The results of other approaches which give predictions for a wide set of the semileptonic 
decay modes are also shown. \\
{\bf The decay \boldmath$D\to\pi,\rho$\unboldmath}. \\
Table \ref{table:ffsd2pi} present the parameters of the fits to the calculated form factors. 
Using the value $V_{cd}=0.22$ \cite{pdg} yields the following decay rates: 
\begin{eqnarray}
\label{d2pi}
\Gamma(D^0\to\pi^-)&=&0.62 \times 10^{10}\,s^{-1}\nonumber \\
\Gamma(D^0\to\rho^-)&=&
\left\{
\begin{array}{lll} 
0.30 \times 10^{10}\,s^{-1},\quad& \Gamma_L/\Gamma_T=1.32   \quad & (LC)\\
0.26 \times 10^{10}\,s^{-1},     & \Gamma_L/\Gamma_T=1.27         & (HQ)
\end{array}
\right.
\end{eqnarray}
Table \ref{table:ratesd2pi} presents the rates for the $HQ$ prescrition of the model and the results of
other approaches versus experimental data. The experimental results are obtained by combining the decay
rates of the $D\to K,K^*$ transition \cite{witherell} with the following ratios 
measured by CLEO \cite{cleopik} 
$\mbox{Br}(D^0\to \pi^-e^+\nu)/\mbox{Br}(D^0\to K^-e^+\nu)=0.103\pm0.039\pm0.013$
and E653 \cite{e653} $\mbox{Br}(D^0\to \rho^-e^+\nu)/\mbox{Br}(D^0\to K^{*-}e^+\nu)=0.088\pm 0.062\pm 0.028$.
The calculated rates seem to be a bit small but nevertheless agree with the experimental 
values within large errors.\\
{\bf The decay \boldmath$D_s\to\eta,\eta',\phi$\unboldmath}. \\
Table \ref{table:ffsds2ss} presents the results on the form factors. The calculated 
decay rates depend on the content of $\eta$ and $\eta'$ mesons and with $|V_{cs}|=0.975$ 
read 
\begin{eqnarray}
\Gamma(D_s\to\eta) &=&0.111\;\sin^2(\varphi)\,ps^{-1} \nonumber\\
\Gamma(D_s\to\eta')&=&0.030\;\cos^2(\varphi)\,ps^{-1}  \nonumber\\
\Gamma(D_s\to \phi)&=&
\left\{
\begin{array}{lll} 
0.047\,ps^{-1},\quad&\Gamma_L/\Gamma_T=1.30    \quad & (LC)\\
 0.040\,ps^{-1},    & \Gamma_L/\Gamma_T=1.28         & (HQ)
\end{array}
\right.
\end{eqnarray}
Here $\varphi=\theta_P+\arcsin(2/\sqrt{6})$ \cite{pdg}. 
The decay rate of $D_s\to \phi$ calculated with the $(HQ)$ prescription agrees well with the results of 
the analysis \cite{isgw2} $\Gamma(D_s\to \phi e^+\nu)=(0.035\pm0.005)ps^{-1}$. 
Table \ref{table:ratesds2ss} compares the results on ratios of branching fractions with 
recent CLEO measurements \cite{cleods} and the ISGW2 model. 
The results for all $\theta_P$ in 
the range $-18^\circ\le\theta_P\le-10^\circ$ compare favourably with the data, 
but the best agreement is observed for $\theta_P=-14^\circ$. \\
{\bf The decay \boldmath$B\to D,D^*$\unboldmath}. \\
This is a very interesting mode as it allows measuring corrections to the HQS
limit. Table \ref{table:ffsb2d} shows the fit parameters of the form factors and
Table \ref{table:iwffs} presents the parameters of the fit to the heavy--quark
form factors in the form 
\begin{equation}
h_i(\omega)=h_i(1)\left[1-\rho_i^2(\omega-1)+\delta(\omega-1)^2\right].
\end{equation}
We find $h_+(1)=0.96$ and $h_-(1)=-0.04$ which compare favourably with the size
of corrections to the HQS limit \cite{neubert}. The values $h_f^{HQ}(1)=0.94$
and $h_f^{LC}(1)=0.9$ both agree with the estimate of Neubert \cite{neubert}
$h_f(1)=0.93\pm0.03$. 
For the ratios of the heavy quark form factors
$R_1=h_g(1)/h_f(1)$ and 
$R_2=h_{a_+}(1)/h_f(1)$ we obtain 
$R^{HQ}_1=1.05$ [$R^{LC}_1=1.1$] and $R^{HQ}_2=0.84$ [$R^{LC}_2=0.88$] 
to be compared with a recent CLEO 
result $R_1=1.18\pm 0.15\pm0.16$ and $R_2=0.71\pm0.22\pm0.07$ and
predictions of the ISGW2 model \cite{isgw2} $R_1=1.27,\;R_2=1.01$, 
Neubert \cite{neubertr} $R_1=1.35,\;R_2=0.79$, and 
Close and Wambach \cite{cw} $R_1=1.15,\;R_2=0.91$. 
CLEO \cite{cleohq} reported the value $|V_{cb}|h_f(1)=0.0351\pm0.0019\pm0.0018\pm0.0008$.
Combining this value with our result $h_f^{HQ}(1)=0.94$ yields 
$$|V_{cb}|=0.0373\pm0.0053\qquad[\mbox{lepton endpoint region in }\bar B\to D^*l\bar\nu].$$
For the decay rates we find
\begin{eqnarray}
\label{b2d}
\Gamma(B\to D)&=&8.712 \times 10^{12}|V_{cb}|^2\,s^{-1} \nonumber\\
\Gamma(B\to D^*)&=&
\left\{
\begin{array}{lll} 
21.0 \times 10^{12}|V_{cb}|^2\,s^{-1},\quad&\Gamma_L/\Gamma_T=1.17    \quad & (LC)\\
23.2\times 10^{12}|V_{cb}|^2\,s^{-1},     &  \Gamma_L/\Gamma_T=1.28        & (HQ)
\end{array}
\right.
\end{eqnarray}
Table \ref{table:ratesb2d} compares the calculated decay rates for the HQ prescription 
with other approaches. Combining our result with a CLEO measurement \cite{cleobd*} 
$\Gamma(\bar B\to D^*l\bar\nu)=[29.9\pm1.9(stat)\pm2.7(syst)\pm2.0(lifetime)]ns^{-1}$ 
yields 
$$|V_{cb}|=0.036\pm0.004\qquad [\mbox{decay rate }\bar B\to D^*l\bar\nu].$$ 
The branching ratio $\mbox{Br}(B^0\to D^-l^+\nu)=(1.9\pm0.5)\%$ and the $B^0$ lifetime 
$\tau_{B^0}=(1.56\pm0.06)\;ps$
\cite{pdg} give the experimental decay rate $\Gamma(B^0\to D^-l^+\nu)=(1.22\pm0.3)10^{10}s^-1$
and comparing with our (HQ) result yields 
$$|V_{cb}|=0.038\pm0.004\qquad [\mbox{decay rate }B^0\to D^-l^+\nu].$$
All the three estimates of $V_{cb}$ agree with each other and the average value is found 
to be 
\begin{equation}
\label{vcb}
|V_{cb}|^{excl}=0.037\pm0.004.
\end{equation}
This is in perfect agreement with the updated values \cite{skwarnicki} 
$|V_{cb}|^{excl}=0.0373\pm0.0045(exp)\pm0.0065(th)$
and $|V_{cb}|^{incl}=0.0398\pm0.0008(exp)\pm0.0004(th).$\\
{\bf The decay \boldmath$B\to\pi,\rho$\unboldmath}. \\
This mode allows a determination of $|V_{ub}|$ and hence reliability of theoretical
predictions is very important. The form factors are given in Table \ref{table:ffsb2pi}.
For the decay rates we find
\begin{eqnarray}
\label{b2pi}
\Gamma(B^0\to\pi^-)&=&7.2 \times 10^{12}|V_{ub}|^2\,s^{-1} \nonumber\\
\Gamma(B^0\to\rho^-)&=&
\left\{
\begin{array}{lll} 
8.44 \times 10^{12}|V_{ub}|^2\,s^{-1},\quad& \Gamma_L/\Gamma_T=0.95  \quad & (LC)\\
9.64 \times 10^{12}|V_{ub}|^2\,s^{-1},     & \Gamma_L/\Gamma_T=1.13        & (HQ)
\end{array}
\right.
\end{eqnarray}
This calculation is in agreement with our previous analysis of this decay mode 
using other sets of the quark model parameters \cite{m1}. 
The calculated decay rates are compared with other theoretical predictions and 
first measurements by CLEO \cite{cleobpi} in Table \ref{table:ratesb2pi}. 
The experimental values are obtained by combining the CLEO results \cite{cleobpi} 
$\mbox{Br}(B^0\to\pi^-l^+\nu)=(1.8\pm0.4\pm0.3\pm0.2)10^{-4}$ and 
$\mbox{Br}(B^0\to\rho^-l^+\nu)=(2.5\pm0.4\pm0.7\pm0.5)10^{-4}$ 
with the $B^0$ lifetime $\tau_{B^0}=(1.56\pm0.06)\;ps$ \cite{pdg}. 
Comparing our results with the experimental values gives
\begin{eqnarray}
|V_{ub}|=0.004\pm0.001\qquad [B\to\pi]\nonumber\\
|V_{ub}|=0.00407\pm0.001\qquad [B\to\rho]\nonumber
\end{eqnarray}
A good agreement of these values with each other shows that we have predicted correctly 
the ratio of the branching fractions $B\to\pi$ and $B\to\rho$. The average value
obtained from the two modes reads 
$$
|V_{ub}|=0.004\pm0.001. 
$$
Taking the $|V_{cb}|$ from (\ref{vcb}) we find 
$$
|V_{ub}/V_{cb}|=0.108\pm0.02
$$
which is perhaps a bit large but nevertheless agrees with the PDG value $|V_{ub}/V_{cb}|=0.08\pm0.02$. 

In conclusion, we have analysed semileptonic decays of heavy mesons within dispersion
formulation of the constituent quark model and found agreement with all available data. 
The extracted values of the CKM matrix elements $V_{cb}$ and $V_{ub}$ are also in 
agreement with estimates of other models. 
Nevertheless, we would like to briefly outline possible sources of
uncertainties in the predictions of the model which should be taken into account in
further analyses:\\
1. We used the parameters of the ISGW2 quark model and a simplified exponential ansatz for the 
wave function. It should be noticed however that the ISGW2 model does not calculate the form
factors through the wave functions; rather a special prescription for constructing the form factors is
formulated. As the analysis of \cite{simula} shows, the form factors of the
heavy--to--light transition can be rather sensitive to the wave function shape. \\
2. We have taken into account only the leading process and neglected the $O(\alpha_s)$ 
corrections. Although the analysis of such corrections in the elastic pion form factor 
at low and high momentum transfers within the LCQM \cite{amn} found only a few $\%$ contribution even 
at $\omega\simeq10-20$ a numerical consideration of this contribution is plausible.\\ 
3. We have neglected the constituent quark transition form factor which has a complicated 
structure at timelike momentum transfers. In particular the quark transition form factor 
should contain a pole at $q^2=M_{res}^2$ with $M_{res}$ 
the mass of a resonance with appropriate quantum numbers.\\
4. We have identified the form factors obtained within the constituent quark model with 
the form factors of the full theory. However, the relationship between these two quantites 
is nontrivial: e.g. the form factors of the full theory acquire logarithmic corrections 
because of renormalization of the quark currents, which are absent in the quark model form
factors. Analysing the $1/m_Q$ expansion Scora and Isgur \cite{isgw2} performed a special
matching procedure for obtaining the form factors of the full theory from their quark--model 
form factors. Although the $1/m_Q$ behaviour of the LCQM form factors studied in 
\cite{mn} is better than that of the generically nonrelativistic form factors in the ISGW model 
\cite{isgw}, the relationship between the LCQM form factors and the form factors of the full theory
should be studied in more detail.

I am grateful to V.Anisovich, I.Narodetskii, K.Ter--Martirosyan, and B.Stech for discussions and 
interest in this work. The work was supported by the Russian Foundation for Basic Researh under 
grant 96--02--18121a. 



\begin{table}
\caption{\label{table:parameters}
Parameters of the quark model.}
\centering
\begin{tabular}{|c||c|c|c|c|c|c|c|c|c|c|c|c|c|c|}
Ref. & $m_u$ & $m_s$ &$m_c$ & $m_b$ &
$\beta_{\pi}$ & $\beta_{K}$ & $\beta_{\eta,\eta'}^{s\bar s}$& 
$\beta_{D}$ & $\beta_{D_s}$ & $\beta_{B}$ &
$\beta_{\rho}$ & $\beta_{K^{*}}$ & $\beta_{\phi}$ & $\beta_{D^*}$  \\
\hline
\hline
ISGW2\cite{isgw2} & 0.33  & 0.55 & 1.82 & 5.2  & 
 0.41  & 0.44 & 0.53  & 0.45  & 0.56  & 0.43  &
 0.30  & 0.33  & 0.37 & 0.38 \\
\end{tabular}
\end{table}

\begin{table}
\caption{\label{table:ffsd2k}
Parameters of the fits to the calculated $D\to K,K^*$ transition form factors.}
\centering
\begin{tabular}{|c|c|c|c|c|c|}

 & {$D \to K$} & \multicolumn{4}{c|}{$D \to K^*$}   \\
\hline
& $f_{+}$ &$g$ &$a_+$  &$f^{LC}$ &$f^{HQ}$ \\
\hline             
$f(0)$                 & 0.781 & 0.28  & $-$0.168 & 1.747  & 1.733  \\
\hline 
$\alpha_1$[GeV$^{-2}$] & 0.201 & 0.24  & 0.189    & 0.0971 & 0.0767 \\
\hline
$\alpha_2$[GeV$^{-4}$] & 0.0086& 0.0135& 0.001    & 0.001  & 0.001  \\
\hline
$f(q^2_{max})$         & 1.2   & 0.35 &$-$0.205   & 1.92   & 1.86 
\end{tabular}
\end{table}

\begin{table}
\caption{\label{table:ratesd2k}
Decay rates for the $D\to K,K^*$ transition in $10^{10}\;s^{-1}$ using $|V_{cs}|=0.975$}
\centering
\begin{tabular}{|c|c|c|c|c|c|c|}
& This work&  WSB \cite{wsb} & ISGW2 \cite{isgw2}& Jaus \cite{jaus} & BBD \cite{bbd} & Exp. \cite{witherell} \\
\hline
$\Gamma(D\to K)$ & $8.7$&$7.56$& $10.0$& $9.6$& $6.5\pm1.3$& $9.0\pm 0.5$ \\
\hline
$\Gamma(D\to K^*)$& $5.38$&$7.73$& $5.4$& $5.5$& $3.7\pm1.2$& $5.1\pm 0.5$  \\
\hline
$\Gamma(K^*)/\Gamma(K)$&  0.62 & 1.02& 0.54 & 0.57 & $0.57\pm0.15$ & $0.57\pm 0.08$   \\
\hline
$\Gamma_L/\Gamma_T$& 1.31& -- & 0.94 & 1.33 & $0.86\pm0.06$ & $1.15\pm0.17$   \\
\end{tabular}
\end{table}

\begin{table}
\caption{\label{table:ffsd2pi}
Parameters of the fits to the calculated $D^0\to\pi^-,\rho^-$ transition form factors.}
\centering
\begin{tabular}{|c|c|c|c|c|c|}
 & {$D \to\pi$} & \multicolumn{4}{c|}{$D \to\rho$}   \\
\hline
& $f_{+}$ &$g$ &$a_+$  &$f^{LC}$ &$f^{HQ}$ \\
\hline             
$f(0)$                 & 0.681& 0.252&$-$0.139 & 1.326& 1.257 \\
\hline
$\alpha_1$[GeV$^{-2}$] & 0.225& 0.274& 0.211   & 0.110& 0.071 \\
\hline
$\alpha_2$[GeV$^{-4}$] & 0.010& 0.017& 0.012   & 0.002& 0.003 \\
\hline
$f(q^2_{max})$         & 1.63  & 0.36 &$-$0.18 & 1.52 & 1.37 
\end{tabular}
\end{table}

\begin{table}
\caption{\label{table:ratesd2pi}
Decay rates for the $D^0\to(\pi^-,\rho^-)e^+\nu$ transition in $10^{10}\;s^{-1}$ using $|V_{cd}|=0.22$. }
\centering
\begin{tabular}{|c|c|c|c|c|c|c|}
& This work&  WSB \cite{wsb} & ISGW2 \cite{isgw2}& Jaus \cite{jaus}  & Ball \cite{ball1} &Exp.  \\
\hline
$\Gamma(D\to\pi)$ &$0.62$& 0.68  & $0.24$& $0.8$&  $0.39\pm 0.08$ & $0.92\pm0.45$ \\
\hline
$\Gamma(D\to\rho)$& $0.26$&$0.67$& $0.12$& $0.33$&  $0.12\pm 0.03$ & $0.45\pm 0.22$  \\
\hline
$\Gamma(\rho)/\Gamma(\pi)$&0.41 &0.98& 0.51 & 0.41 &  $0.3\pm0.1$  & $0.5\pm 0.35$   \\
\hline
$\Gamma_L/\Gamma_T$&1.27 &0.91 & 0.67 & 1.22 &  $1.31\pm0.11$ & --  \\
\end{tabular}
\end{table}

\begin{table}
\caption{\label{table:ffsds2ss}
Parameters of the fits to the calculated $D_s\to s\bar s,\phi$ transition form factors.}
\centering
\begin{tabular}{|c|c|c|c|c|c|}
 & {$D_s \to s\bar s$} & \multicolumn{4}{c|}{$D_s \to\phi$}   \\
\hline
                       &$f_{+}$ &  $g$  &$a_+$&$f^{LC}$&$f^{HQ}$\\
\hline             
$f(0)$                 & 0.800  & 0.266&$-$0.149&  1.806& 1.710  \\
\hline
$\alpha_1$[GeV$^{-2}$] & 0.192  & 0.246&   0.201&  0.111& 0.085 \\
\hline
$\alpha_2$[GeV$^{-4}$] & 0.008  & 0.015&   0.009&$-$.003& 0.013
\end{tabular}
\end{table}

\begin{table}
\caption{\label{table:ratesds2ss}
Ratio of the decay rates for the $D_s\to\eta,\eta',\phi$ transitions.}
\centering
\begin{tabular}{|c|c|c|c|c|c|c|}
& \multicolumn{3}{c|}{This work} & \multicolumn{2}{c|}{ISGW2 \cite{isgw2}} & Exp. \cite{cleods}\\
\hline
& $\theta_P=-10^\circ$  & $\theta_P=-14^\circ$ & $\theta_P=-20^\circ$ 
& $\theta_P=-10^\circ$  & $\theta_P=-20^\circ$ &      \\ 
\hline
$\Gamma(\eta)/\Gamma(\phi)$ & 1.45 & 1.24 & 0.9 & 1.2 & 0.8 & $1.24\pm 0.27$\\
\hline
$\Gamma(\eta')/\Gamma(\phi)$& 0.4  & 0.46 & 0.6 & 0.5 & 0.7 & $0.43\pm 0.18$\\
\hline
$\Gamma(\eta')/\Gamma(\eta)$& 0.27 & 0.37 & 0.67&     &     & $0.35\pm 0.16$
\end{tabular}
\end{table}


\begin{table}
\caption{\label{table:ffsb2d}
Parameters of the fits to the calculated $B\to D,D^*$ transition form factors.}
\centering
\begin{tabular}{|c|c|c|c|c|c|c|}
 & \multicolumn{2}{c|}{$B \to D$} & \multicolumn{4}{c|}{$B \to D^*$}   \\
\hline
& $f_{+}$ & $f_{-}$ &$g$ &$a_+$ &$f^{LC}$ &$f^{HQ}$ \\
\hline             
$f(0)$                 & 0.684  &$-$0.337& 0.093  & $-$0.0764 & 4.533   & 4.729   \\
\hline
$\alpha_1$[GeV$^{-2}$] & 0.0386 & 0.039  & 0.0416 & 0.0387    & 0.02193 & 0.02195 \\
\hline
$\alpha_2$[GeV$^{-4}$] & 0.00042& 0.00038& 0.00048& 0.00043   & 0.00007 & 0.00007 \\
\hline
$f(q^2_{max})$         & 1.12   &$-$0.56 & 0.15   &$-$0.12    & 5.85    & 6.10 
\end{tabular}
\end{table}

\begin{table}
\caption{\label{table:iwffs}
Parameters of the heavy--quark form factors for the $B\to D,D^*$ transition.}
\centering
\begin{tabular}{|c|c|c|c|c|c|c|c|}

&\multicolumn{2}{c|} {$B \to D$} & \multicolumn{5}{c|}{$B \to D^*$}   \\
\hline
& $h_{+}$ &$h_-$ &$h_g$ &$h_{a_+}$ &$ h_{a_-}$ & $h_f^{LC}$ & $h_f^{HQ}$ \\
\hline             
$h(1)$  & 0.96 & $-$0.04 & 0.99 & 0.79     & 0.92  & 0.90 & 0.94  \\
\hline
$\rho^2$& 0.91 & $-$     & 1.04 & 1.20     & 1.08  & 1.10 & 1.06  \\
\hline
$\delta$& 0.37 & $-$     & 0.54 & 0.63     & 0.50  & 0.55 & 0.53  
\end{tabular}
\end{table}

\begin{table}
\caption{\label{table:ratesb2d}
Decay rates for the $B\to D,D^*$ transition in $ps^{-1}$ . }
\centering
\begin{tabular}{|c|c|c|c|c|c|}
& This work &  WSB \cite{wsb} & ISGW2 \cite{isgw2}& Jaus \cite{jaus} & Exp. \\
\hline
$\Gamma(B\to D)$ & $ 8.7|V_{cb}|^2$&
$ 8.1|V_{cb}|^2$&
$11.9|V_{cb}|^2$&
$ 9.6|V_{cb}|^2$&
$1.27\pm 0.3\times10^{-2}$ \cite{pdg}\\
\hline
$\Gamma(B\to D^*)$&$23.2|V_{cb}|^2$&
$21.9|V_{cb}|^2$&
$24.8|V_{cb}|^2$&
$25.3|V_{cb}|^2$&
$2.99\pm 0.66\times10^{-2}$ \cite{cleobd*}\\
\hline
$\Gamma(D^*)/\Gamma(D)$&2.65 &
2.71& 2.08 & 2.64 &  $2.35\pm 1.3$   \\
\hline
$\Gamma_L/\Gamma_T$&1.28 &
-- & 1.04 & -- & $1.24\pm0.16$ \cite{cleogr}   \\
&   &      &    &      & $0.85\pm0.45$ \cite{argusgr}  \\
\end{tabular}
\end{table}
\begin{table}
\caption{\label{table:ffsb2pi}
Parameters of the fits to the calculated $\bar B^0\to\pi^+,\rho^+$ transition form factors.}
\centering
\begin{tabular}{|c|c|c|c|c|c|}
 & {$B \to\pi$} & \multicolumn{4}{c|}{$B \to\rho$}   \\
\hline
                       &$f_{+}$ &  $g$  &$a_+$&$f^{LC}$&$f^{HQ}$\\
\hline             
$f(0)$                 & 0.2927 & 0.0356&$-$0.0256& 1.025  & 1.098  \\
\hline
$\alpha_1$[GeV$^{-2}$] & 0.0511 & 0.0635& 0.0567  & 0.032  & 0.0316 \\
\hline
$\alpha_2$[GeV$^{-4}$] & 0.00068& 0.0012& 0.0010  & 0.00028& 0.00038\\
\hline
$f(q^2_{max})$         & 2.30   & 0.17  &$-$0.097 & 2.19   & 2.12 
\end{tabular}
\end{table}

\begin{table}
\caption{\label{table:ratesb2pi}
Decay rates for the $\bar B^0\to(\pi^+,\rho^+)e\bar\nu$ transition in $ps^{-1}$.}
\centering
\begin{tabular}{|c|c|c|c|c|c|c|}
& This work&  WSB \cite{wsb} & ISGW2 \cite{isgw2}& Jaus \cite{jaus} & Ball \cite{ball1,ball2}&
Exp.\cite{cleobpi} \\
\hline
$\Gamma(B\to\pi)$ & $7.2|V_{ub}|^2$& 
$7.4|V_{ub}|^2$& 
$9.6|V_{ub}|^2$& 
$10.0|V_{ub}|^2$& 
$5.1\pm1.1|V_{ub}|^2$& 
$1.2\pm 0.6\times 10^{-4}$ \\
\hline
$\Gamma(B\to\rho)$&$9.64|V_{ub}|^2$& 
$26.0|V_{ub}|^2$& 
$14.2|V_{ub}|^2$& 
$19.1|V_{ub}|^2$& 
$12\pm4(14.5\pm4.5)|V_{ub}|^2$& 
$1.67\pm 1.0 \times 10^{-4}$ \\
\hline
$\Gamma(\rho)/\Gamma(\pi)$&1.34 &
3.5& 1.48 & 1.91 & $2.35\pm1.2$ &$-$   \\
\hline
$\Gamma_L/\Gamma_T$&1.13 &
1.34& 0.3  & 0.82 & $0.06\pm0.02$ ($0.52\pm0.1$) & $-$  
\end{tabular}
\end{table}

\begin{figure}[1]
\begin{center}  
\mbox{\epsfig{file=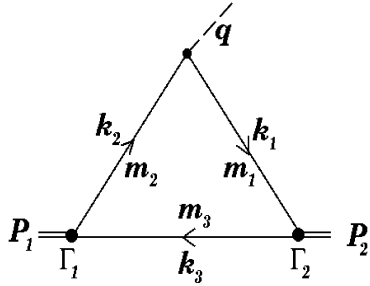,height=4.cm}  }
\end{center}
\caption{One-loop graph for a meson decay.\label{fig:1}}
\end{figure}

\end{document}